\documentclass[twocolumn,preprintnumbers,amsmath,amssymb,superscriptaddress,prl,floatfix]{revtex4-1}
\usepackage{graphicx}
\usepackage{bm}
\usepackage{xcolor}

\begin{document}

\title{Evidence for bond-disproportionation in LiNiO$_2$ from x-ray absorption spectroscopy}


\author{R.  J. Green}
\affiliation{Stewart Blusson Quantum Matter Institute, University of British Columbia, Vancouver, British Columbia, V6T 1Z4, Canada}
\affiliation{Department of Physics and Engineering Physics, University of Saskatchewan, Saskatoon, Saskatchewan S7N 5E2, Canada}

\author{H. Wadati}
\affiliation{Department of Physics and Astronomy, University of British Columbia, Vancouver, British Columbia, V6T 1Z1, Canada}

\author{T. Z. Regier}
\affiliation{Canadian Light Source, University of Saskatchewan, Saskatoon, Saskatchewan, S7N 2V3, Canada}

\author{A. J. Achkar}
\affiliation{Department of Physics and Astronomy, University of Waterloo, Waterloo, Ontario, N2L 2C3, Canada}

\author{C. McMahon}
\affiliation{Department of Physics and Astronomy, University of Waterloo, Waterloo, Ontario, N2L 2C3, Canada}

\author{J. P. Clancy}
\affiliation{Department of Physics and Astronomy, McMaster University, Hamilton, Ontario, L8S 4M1, Canada}

\author{H. A. Dabkowska}
\affiliation{Brockhouse Institute of Materials Research, McMaster University, Hamilton, Ontario, L8S 4M1, Canada}

\author{B. D. Gaulin}
\affiliation{Department of Physics and Astronomy, McMaster University, Hamilton, Ontario, L8S 4M1, Canada}
\affiliation{CIFAR, 661 University Ave., Toronto, Ontario, M5G 1M1, Canada}

\author{G. A. Sawatzky}
\affiliation{Stewart Blusson Quantum Matter Institute, University of British Columbia, Vancouver, British Columbia, V6T 1Z4, Canada}
\affiliation{Department of Physics and Astronomy, University of British Columbia, Vancouver, British Columbia, V6T 1Z1, Canada}

\author{D. G. Hawthorn}
\affiliation{Department of Physics and Astronomy, University of Waterloo, Waterloo, Ontario, N2L 2C3, Canada}


\begin{abstract}
The electronic structure of LiNiO$_2$, a promising Li-ion battery cathode material, has remained a challenge to understand due to its highly covalent yet correlated nature. Here we elucidate the electronic structure in LiNiO$_2$ and the related compound NaNiO$_2$ using x-ray absorption spectra (XAS) and quantum many-body calculations. Notably, we use inverse partial fluorescence yield to correctly measure the Ni $L$-edge XAS, which is inaccurate using conventional methods. We show that the XAS are indicative of a strong Jahn-Teller effect in NaNiO$_2$ and a bond disproportionated state in LiNiO$_2$, supporting a theory of a high-entropy, glassy disproportionated state that stabilizes charging cycles in LiNiO$_2$.
\end{abstract}



\maketitle

LiNiO$_2$ is most well known as a potential cathode material for Li-ion batteries.\cite{Liu2015}  However, despite extensive study, the microscopic electronic structure of LiNiO$_2$ remains unresolved.  LiNiO$_2$ and its relative NaNiO$_2$ both consist of layers of edge-sharing NiO$_6$ octahedra arranged on a triangular lattice with Li or Na residing in-between the NiO$_2$ layers.  In these compounds Ni is in a formally Ni$^{3+}$ 3$d^7$ configuration, with a large octahedral crystal/ligand field yielding a low-spin $t_{2g}^6e_g^1$ configuration.  Since the $t_{2g}^6e_g^1$ configuration is susceptible to a Jahn-Teller (JT) distortion, it is expected that the octahedra are distorted in a manner that lifts the degeneracy of the $e_g$ states. In  NaNiO$_2$  a cooperative Jahn-Teller distortion indeed occurs, resulting in both ferro-orbital ordering of $d_{3z^2-r^2}$ states below 460 K and eventual anti-ferromagnetic ordering of spins.\cite{Borgers1966,Mostovoy2002,Lewis2005,Meskine2005}  In contrast, LiNiO$_2$ does not exhibit long-range order to the lowest observable temperatures.\cite{Arai1995,Chung2005} Rather, EXAFS and neutron pair distribution function measurements have shown evidence for some local Jahn-Teller distortions of NiO$_6$ octahedra, without long-range ordering of the distortions.\cite{Rougier1995,Chung2005} 

The reason for the lack of ordering in LiNiO$_2$ is unresolved. LiNiO$_2$ may be an example of an $S = 1/2$ antiferromagnet on a triangular lattice, which is a candidate for a frustrated spin liquid. However, LiNiO$_2$ also exhibits Li:Ni cross-substitution, with Li on the Ni site, and vice-versa, that may introduce sufficient disorder to inhibit long range JT and anti-ferromagnetic order.\cite{Petit2006} In addition, recent work has indicated that LiNiO$_2$ may be a charge and bond disproportionated glass.\cite{Chen2011,Foyevtsova2019}  Density functional theory calculations reveal a large number of nearly degenerate ground states that involve local Jahn Teller distortions and both bond and charge disproportionation.\cite{Foyevtsova2019}  The bond disproportionations in such a state are similar to those found in the perovskite rare-earth nickelates,  $R$NiO$_3$.\cite{Mizokawa2000,Mazin2007,Park2012,Johnston2014,Lao2013,Subedi2015,Green2016} 

Like NaNiO$_2$ and LiNiO$_2$, a key feature of  $R$NiO$_3$  is the high 3+ formal oxidation state of Ni.  The Ni$^{3+}$ is uncommon, with the electronegativity of Ni typically yielding a valence closer to $2+$, as in NiO.  In  $R$NiO$_3$,  it is proposed that this high oxidation state  leads to these  compounds having a negative charge-transfer energy\cite{Green2016,Sawatzky2016} in the Zaanan-Sawatzky-Allen (ZSA) classification scheme.\cite{Mizokawa1995,Zaanen1985} In this model, the actual Ni valence is close to 2+, with the excess holes present on the neighbouring oxygen. Evidence for the negative charge transfer energy has come from  resonant inelastic x-ray scattering.\cite{Bisogni2016} Additionally, a successful description of both the XAS and resonant x-ray scattering in $R$NiO$_3$ is achieved in a model that includes hybridization between neighbouring NiO$_6$ clusters and bond-disproportionation, wherein the hybridization of O states on one NiO$_6$ differs from that of its neighbour.\cite{Green2016}

Evidence that LiNiO$_2$ is also a negative-charge-transfer system, similar to  $R$NiO$_3$,  has come from x-ray absorption measurements at the O $K$ edge in Li$_{x}$Ni$_{2-x}$O$_2$.  These measurements show a large oxygen pre-peak that grows with Li content, indicative of a large hole content in O $2p$ states,\cite{Kuiper1989} reminiscent of hole doped cuprates.\cite{Chen1991}  Moreover, the  spectrum  of NaNiO$_2$ \cite{VanVeenendaal1994} appears similar to that of  $R$NiO$_3$.\cite{Piamonteze2005,Bodenthin2011,Freeland2016}  However, measurements of the Ni $L$ edge of LiNiO$_2$ have often indicated a local electronic configuration of Ni similar Ni$^{2+}$ in NiO or an admixture of Ni$^{2+}$ and Ni$^{3+}$ rather than negative charge transfer Ni$^{3+}$,\cite{Abbate1991,Montoro1999,Kang2007} leading to uncertainty in the overall electronic structure of this compound.


One of the challenges of interpreting the XAS of NaNiO$_2$ and LiNiO$_2$ (as well as AgNiO$_2$)\cite{Kang2007} is that the x-ray absorption can be difficult to measure correctly. Both NaNiO$_2$ and LiNiO$_2$ are hydroscopic, exhibiting Ni$^{2+}$ on their surfaces after exposure to air. Accordingly, surface sensitive total electron yield (TEY) measurements of XAS, having a probing depth of order 50 \AA, may inadvertently probe a contaminated surface that forms a more stable oxide and is not characteristic of the bulk. To circumvent this surface sensitivity, bulk  sensitive total fluorescence yield (TFY) measurements of XAS may be employed. However, at the Ni $L$ edge in systems such as LiNiO$_2$ where Ni makes the largest contribution to the total absorption cross-section, TFY spectra are heavily distorted by self-absorption effects\cite{Eisebitt1993,Achkar2011a,Achkar2011b}  and inherent non-linearities in the scattering process.\cite{DeGroot1994,Kurian2012,Green2014}  In contrast, since O is a smaller contributor to the total absorption co-efficient, TFY measurements at the O $K$ edge remain reasonably accurate. Here we utilize inverse partial fluorescence yield spectroscopy (IPFY)\cite{Achkar2011a,Achkar2011b} to provide a bulk sensitive and  distortion  free measurement of XAS at the Ni $L$ edge.

\begin{figure}
\includegraphics[width=\linewidth]{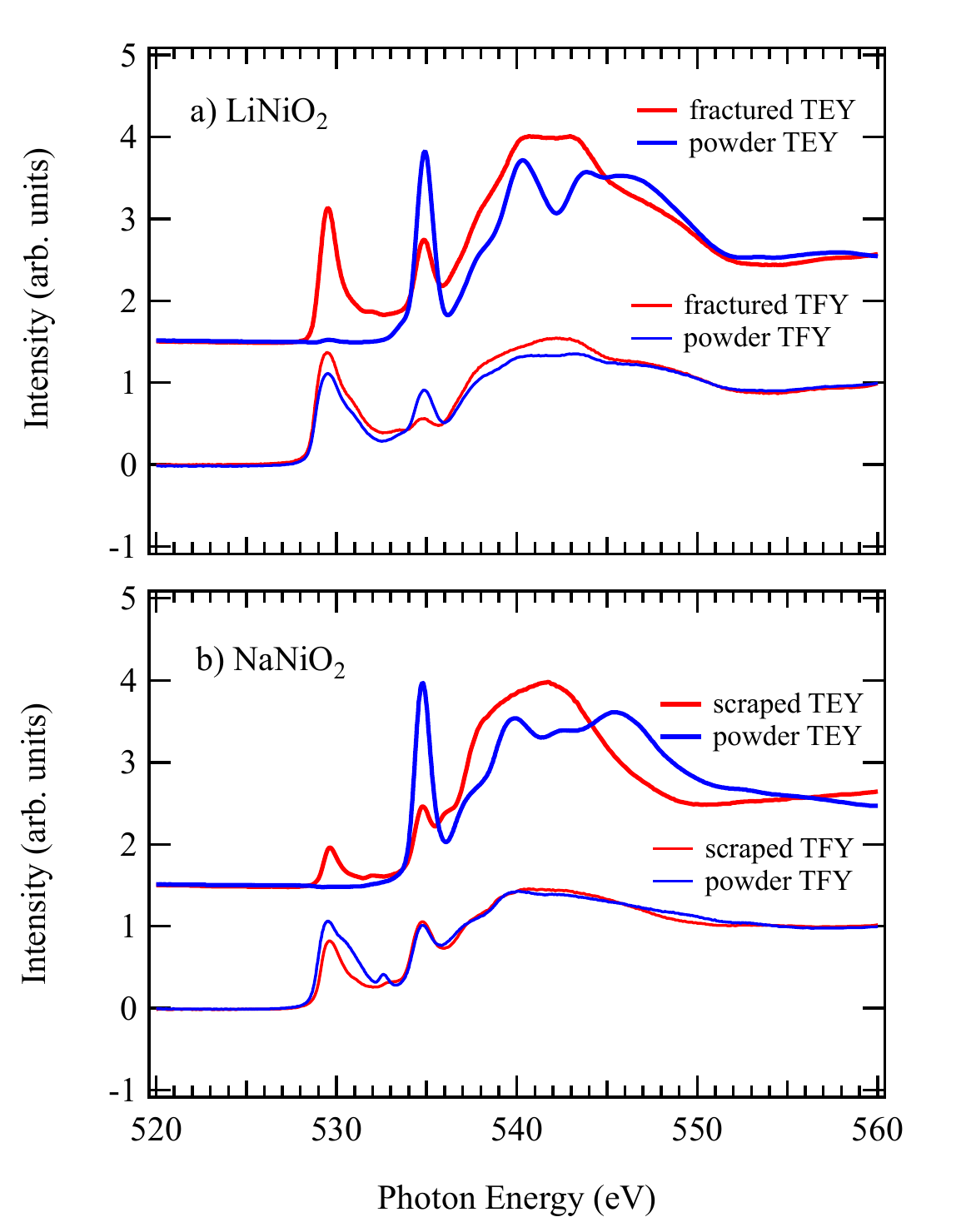}
\caption{X-ray absorption measurements at the O $K$ edge in a) LiNiO$_2$ and b) NaNiO$_2$.  Significant differences in TEY measurements are observed between powder samples exposed to air vs.\ samples that are scraped or fractured in-vacuum.  Samples that are fractured or scraped in vacuum reveal a large O pre-peak at 529.5 eV in both TEY and bulk sensitive TFY.}
\label{fig1OK}
\end{figure}

X-ray absorption measurements were performed on polycrystalline powder samples of LiNiO$_2$ and NaNiO$_2$ at the SGM beamline at the Canadian Light Source.  The samples were prepared with different surface treatments, either as powder samples that have been exposed to air and pressed onto carbon tape or sintered powders that are either fractured or scraped in vacuum to reveal material that has not been previously exposed to air.  For LiNiO$_2$ the  latter  procedure produced samples with less surface contamination.  However, for NaNiO$_2$, sintering appeared to change the bulk of the sample relative to the initial powders.  TEY measurements were performed by measuring the sample drain current.  TFY and IPFY were measured using an energy resolved silicon drift detector.\cite{Achkar2011a,Achkar2011b}  IPFY uses the O $K_{\alpha}$ x-ray emission, the inverse of which has been shown to be proportional to the x-ray absorption co-efficient plus an approximately energy independent offset, providing a measure of the absorption co-efficient that is more bulk sensitive than TEY, like TFY, but does not suffer from self-absorption effects that can be prominent in concentrated systems such as LiNiO$_2$.\cite{Achkar2011a,Achkar2011b}

Figure~\ref{fig1OK} presents  O $K$ edge measurements of LiNiO$_2$ and NaNiO$_2$.  Similar to previous measurements,\cite{Kuiper1989} a large pre-peak is observed (at 529.5 eV),  indicative of a large hole content in O 2$p$ states.  Similar intensities of the O prepeak are observed in both  NaNiO$_2$  and LiNiO$_2$.  Note these measurements also indicate the role of surface oxidation in LiNiO$_2$.  Surface sensitive TEY measurements exhibit a pronounced prepeak in a sample that is sintered and fractured in vacuum, whereas a powder sample exposed to air exhibits a negligible prepeak at 529 eV and a larger peak at 535 eV, which we associate with surface  reduction. Kuiper et.\ al.\ associate this 535 eV peak with Li$_2$O.\cite{Kuiper1989}  We note, however, that a peak at approximately the same energy is also observed in the TEY of NaNiO$_2$ powder, suggesting an alternate  origin.   

\begin{figure}
\includegraphics[width=\linewidth]{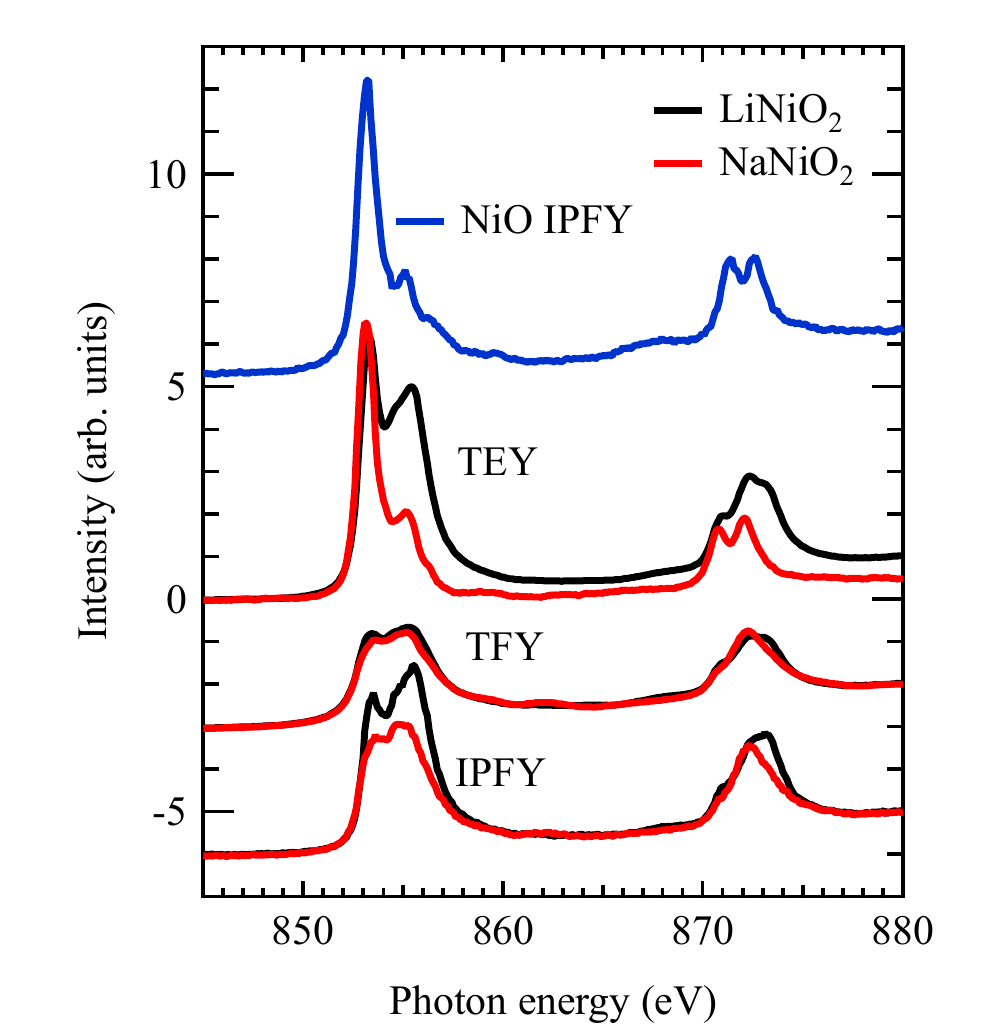}
\caption{X-ray absorption measurements at the Ni $L$ edge in LiNiO$_2$ and NaNiO$_2$ comparing TEY, TFY and IPFY measurements.  For comparison, XAS in NiO from ref.~\cite{Achkar2011b} is also shown.}
\label{fig2LiNiO2}
\end{figure}

Measurements of the Ni $L$ edge are shown in Fig. \ref{fig2LiNiO2}. For LiNiO$_2$, the surface sensitive TEY  spectrum is  similar to that of NiO. The NaNiO$_2$  TEY spectrum also exhibits  a prominent peak at 853 eV indicative of Ni$^{2+}$ on the  surface,  although the effect is less dramatic than LiNiO$_2$.  In contrast, the more bulk sensitive TFY and IPFY measurements do not exhibit a NiO like lineshape.  The IPFY and TFY spectra have similar lineshapes. However, the TFY spectra suffer from a prominent self-absorption effect that reduces the intensity of the $L_3$ peak relative to the $L_2$ peak,  limiting possible comparisons between materials as well as with theory.   The IPFY spectra, which  do  not suffer from self-absorption effects, are expected to provide a more accurate measure of the energy dependence of the absorption co-efficient.

\begin{figure}
\def\svgwidth{\linewidth}
\includegraphics[width=\linewidth]{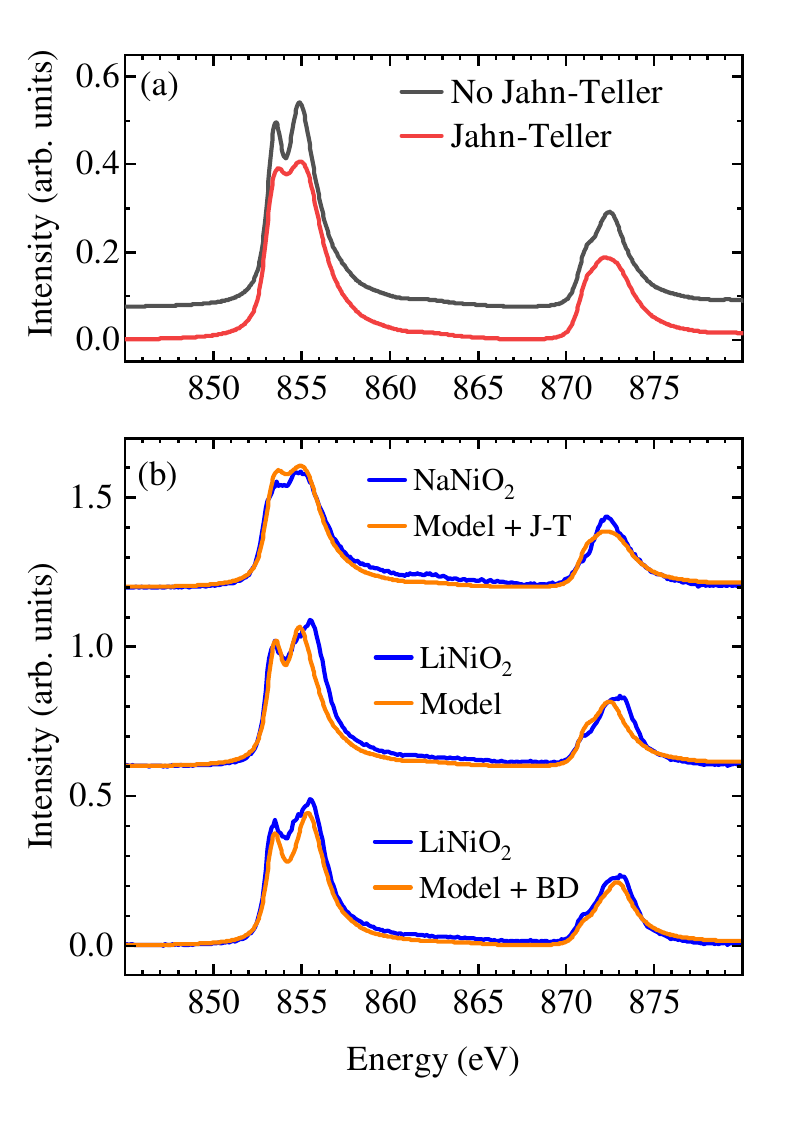}
\caption{Calculation of the Ni $L$ edge x-ray absorption using a full-correlation, double-cluster model.\cite{Green2016} (a) The effect of the Jahn-Teller distortion in the simulations is shown. (b) The model with Jahn-Teller (J-T) is compared to experimental data for NaNiO$_2$. The model without (middle) and with (lower) bond disproportionation (BD) is compared to experimental data for LiNiO$_2$.  }
\label{fig4calc}
\end{figure}

Coarsely, the resulting the Ni $L$-edge spectra are similar for NaNiO$_2$ and LiNiO$_2$ and, despite having edge sharing instead of corner sharing octahedra, are also similar to that of  $R$NiO$_3$ compounds.\cite{Piamonteze2005,Bodenthin2011,Freeland2016}  However, the detailed lineshape of the spectra exhibit key differences for NaNiO$_2$ and LiNiO$_2$ that are signatures of Jahn Teller distortions and bond disproportionation.  The latter exhibits a more pronounced peak splitting at the $L_3$ edge and a more distinct shoulder on the low energy side of the $L_2$ edge. These features have been shown to arise due to strong Ni-Ni interactions in these covalent materials and are enhanced by bond disproportionation.\cite{Green2016} Adapting the full-multiplet, nickelate double cluster model of Ref. \onlinecite{Green2016} to the present case, we can examine the differences in the XAS of NaNiO$_2$ and LiNiO$_2$ in more detail. In figure~\ref{fig4calc}(a) we show calculations of the Ni $L$ edge XAS using the double cluster model, with and without a tetragonal Jahn-Teller effect.\footnote{For the Jahn-Teller case, our hopping integrals are $V_{3z^2-r^2}=2.44 \mathrm{~eV}$, $V_{x^2-y^2}=3.25 \mathrm{~eV}$, $V_{xy}=1.42 \mathrm{~eV}$, $V_{xz}=V_{yz}=1.89 \mathrm{~eV}$. All other parameters are the same as in Ref. \onlinecite{Green2016}, unless otherwise noted in the text. } The Jahn-Teller effect is introduced via tetragonal distortions to the Ni-O hopping integrals using Harrison's formulae.\cite{Harrison1983} The introduction of the Jahn-Teller effect captures the main differences in the spectra of NaNiO$_2$ and LiNiO$_2$.  In the absence of Jahn-Teller, the $L_3$ edge exhibits a robust peak splitting and the $L_2$ edge a pronounced low energy shoulder, similar to the spectrum of LiNiO$_2$.  Upon the introduction of the Jahn-Teller effect these features are suppressed, similar to NaNiO$_2$.

In Fig. \ref{fig4calc}(b) we present a closer comparison of the model calculations and background-subtracted experimental XAS spectra.  The calculation with the Jahn-Teller effect is shown to agree remarkably well with that of NaNiO$_2$.  However, for the calculation without the Jahn-Teller effect, there are some slight discrepencies with the spectrum of LiNiO$_2$. In particular, the peak splitting is stronger in experiment.  As discussed above, this peak splitting is driven in part by Ni-Ni intersite interactions and can be enhanced by bond disproportionation.\cite{Green2016} Given the similar structures of the two compounds, and the agreement of the calculation with the spectrum of NaNiO$_2$, an appropriate level of Ni-Ni intersite interaction is expected to be present in the calculation for LiNiO$_2$.  This suggests that the slight disagreement is due to the presence of bond disproportionation in the experiment that must be accounted for in the model.  Upon the inclusion of bond disproportionation (in a manner shown to be reliable for the perovskite nickelates \cite{Green2016} but with a slightly larger value of $\delta d=0.06 \mathrm{~\AA}$), the agreement between calculation and experiment improves drastically, as shown in Fig. \ref{fig4calc}(b).  This result provides strong support for the existence of bond disproportionation in LiNiO$_2$, as recently theorized.\cite{Foyevtsova2019}

Our IPFY data of NaNiO$_2$ and LiNiO$_2$ provide some of the most accurate XAS measurements of these compounds to date.  The strong similarity of the spectra to those of perovskite nickelates shows the present materials to be negative charge transfer compounds. Using a quantum many body double cluster model, we confirm the existence of a global Jahn-Teller distortion in NaNiO$_2$, and find evidence for a lack of significant Jahn-Teller effects in LiNiO$_2$. Instead, the calculation results provide evidence for a significant degree of bond disproportionation in LiNiO$_2$.  Such a finding provides support for recent theoretical work predicting the existence of a high-entropy, electronic glassy-like disproportionated state in LiNiO$_2$ which is responsible for its strong stability under repeated charge/discharge cycles.

Acknowledgments: This work was supported by the Natural Sciences and Engineering Research Council of Canada (NSERC), the Stewart Blusson Quantum Matter Institute and the Canada First Research Excellence Fund, Quantum Materials and Future Technologies and Transformative Quantum Technologies Programs. Research described in this paper was performed at the Canadian Light Source, which is funded by the CFI, the NSERC, the National Research Council Canada, the Canadian Institutes of Health Research, the Government of Saskatchewan, Western Economic Diversification Canada, and the University of Saskatchewan.  B.D.G. acknowledges support from the CIFAR as a CIFAR Fellow.
\bibliography{LiNiO2XAS}

\end{document}